# CSIEC (Computer Simulator in Educational Communication): An Intelligent Web-Based Teaching System for Foreign Language Learning

Jiyou Jia
University of Augsburg
Germany
jiyou.jia@student.uni-augsburg.de

**Abstract**: In this paper we present an innovative intelligent web-based computer-aided instruction system for foreign language learning: CSIEC (Computer Simulator in Educational Communication). This system can not only grammatically understand the sentences in English given from the users via Internet, but also reasonably and individually speak with the users. At first the related works in this research field are analyzed. Then we introduce the system goals and the system framework, i.e., the natural language understanding mechanism (NLML, NLOMJ and NLDB) and the communicational response (CR). Finally we give the syntactic and semantic content of this instruction system, i.e. some important notations of English grammar used in it and their relations with the NLOMJ.

## Introduction

We have conducted an experiment of the application of a keywords-based human-computer dialog system with natural language (chatbot) on the teaching of foreign languages (Jia, 2004). Findings about the dialogs between the user and the computer indicate that the dialogs are mostly very short because the user finds the computer are much less intelligent as a human and the responses from the computer are mostly repeated and irrelevant with the topics and context of the dialog. But this study indicates also that many participants in this experiment are very interested in this on-line system which can function as a chatting partner speaking the foreign language. The reasons are: this system is accessible anywhere at anytime on contrary to that it is not easy to find a human chatting partner speaking this language as the mother language, and the learners are more confident confronting with a robot or a computer program which is obviously less intelligent as the human themselves. It would be pedagogically attractive for the learners to chat with such a system of artificial intelligence which could "really" understand the natural language and reasonably generate the natural language to form a human-like dialog.

With this motivation we have designed an innovative web-based human-computer communication system with natural language: CSIEC (Computer Simulator in Educational Communication). The system is still under development and the test version is freely accessible at http:///www.csiec.de or http://www.csiec.com.

## Related Works

Natural language processing (NLP) is a very important research field in artificial intelligence and computer linguistics. In the past 50 years great progress has been made in this field, as a vast of references have pointed out (see Allen, 1995; Dale etc., 2000; Iwanska, etc., 2000). The milestone in the history of modern linguistics is Chomsky's concept of generative grammar for natural language with the efforts of describing all possible sentences (Chomsky, 1956, 1965, 1969, 1988). Symbolic approaches to NLP have their origins in generative linguistics. Ideas from linguistic theory, particularly relative to syntactic description along the line proposed by Chomsky and others, were absorbed by researchers in the field, and refashioned in various ways to play a role in working computational systems; in many ways, research in linguistics and the philosophy of the language set the agenda for explorations in NLP. Chomsky-grammar describes the natural language with a rule system, which can be illustrated by an example (Chomsky, 1969, P. 57) shown in Figure 1. The grammar with this rule generates the string "*John saw Bill*" with the phrase-marker shown in Figure 2.

```
S→ NP VP
VP→ V NP
NP→ John, Bill
V→ saw
```

Fig. 1 An example of the rule system of Chomsky-grammar

```
        S
       / \
      /   \
     NP   VP
     |    / \
   John  V NP
         |  |
       saw Bill
```

Fig. 2 An example of phrase marker

The actual English grammar is much more complicated than this example and therefore a corresponding complicated rules system is needed to the English grammar. However work in syntactic description has always been the most thoroughly detailed and worked-out aspect of linguistic inquiry, so that at this level a great deal has been borrowed by NLP researchers. There has been a vast literature on syntactic formalisms and parsing algorithms based on the notation of Chomsky-grammar, like AGFL-project (Affix Grammars over a Finite Lattice) from University of Nijmegen (Koster, 1991), FDG (Functional Dependency Grammar) language analysis system from University of Helsinki (Tapanainen & Järvinen, 1997), Transition network grammars and parsers described by Woods (Woods, 1970, 1973), Chart-based parsers described by Kay (Kay, 1973) and Horn-clause-based parsers described and compared with transition network systems by Pereira and Warren (Pereira and Warren, 1980), etc. So with these parsers and a meticulous describing of the grammar in a natural language the sentences can be well decomposed into their grammatical elements, as the above example. But the problem is that only we, human being, know the meaning of the notations in the parsing results such as NP(nounpart), VP(Verbpart), subject, object, etc., but the computer program doesn't recognize them as the elements in the sentences and therefore can do nothing further with the parsing result.

Joseph Weizenbaum programmed ELIZA, a program operating within the MAC time-sharing system at MIT which makes certain kinds of natural language conversation between man and computer possible. Input sentences are analyzed on the basis of decomposition rules which are triggered by key words appearing in the input text. Responses are generated by reassembly rules associated with selected decomposition rules (Weizenbaum, 1966). The program applied pattern matching rules to the human's statements to figure out its replies. This keywords-based mechanism or pattern-matching mechanism is widely used in the chatbot systems until today. For example, the ALICEBOT used in our experiment is also such a kind of systems (Jia, 2004). But this mechanism has some inevitable shortcomings (Jia, 2003). It grasps only the meanings of the keywords in the sentences, but not the meaning of the whole sentences. In order to make a reasonable response to the whole sentence we must use the whole sentence as the keywords. But in fact we can't put all the possible sentences occurring in the human conversations into the database of the keywords. So the generative nature of human natural language, i.e., to produce unlimited sentences from limited vocabulary, can not be reflected at all by the pattern-matching mechanism. As we have pointed out (Jia, 2003), the muster-matching mechanism is only fit for the idioms and polite formulas such as "*Hello!*", "*how are you*?", for which the syntax analysis is superfluous and even useless, and a response can be directly given without thinking.

Terry Winograd made an important contribute to natural language understanding in his project "SHRDLU" (Winograd, 1972). The SHRDLU program can be viewed historically as one of the classic examples of how difficult it is for a programmer to build up a computer's semantic memory by hand. In making the LISP program Winograd was concerned with the problem of providing a computer with enough "understanding" to be able to use natural language. He restricted the program's intellectual world to a simulated "world of toy blocks". The program could accept commands such as, "*Move the blue block*," and carry out the requested action using a simulated block-moving arm. The program could also respond verbally, for example, "*I do not know which blue block you mean*." It dealt in an integrated way with all the aspects of language: syntax, semantics, inference, and knowledge and reasoning about the subject it describes, i.e. domain knowledge. The dialogs history are saved and referenced for the later answers to the user. These successful thoughts should be inherited in the design of human-computer interaction (HCI) systems with natural language.

## System Goals

As we have introduced above, our ideal human-computer communication system should function as a chatting partner for foreign language, that is, it should be able to chat with the user in the given language, e.g. English, like a human partner as real as possible who has his(her) personality and emotion. The conversations are not restricted to a given topic.

We select English as the conversation language because it is the most popular foreign language and in Internet there are freely downloadable materials about English like lexicon, grammar parser, semantics networks for non-commercial use.

The ideal input method from the user should be acoustic, i.e. a speech recognition system should be used to convert the speech into texts. But regarding the state of the art of the speech recognition technology, we use only the input method via the keyboard.

The ideal output method for the computer system should not only be text, but also be acoustic, i.e. a speech synthesis system should be used to convert the output text to speech.

Additionally this system should still use the Server/Client modal to work as a chatting server which can be accessed by multi-users synchronically.

Comprehensively we try to create an interactive "live" environment for the learners of English as a foreign language where they could chat with a "partner" in English about any thing and at any time, what is lacked in most of the current CAI-systems for foreign language learning.

## System Architecture

Summarizing the historical experiences in designing such systems and considering our system goals, we explore a human-computer communication system (besides the speech recognition and speech synthesis) including the following components:

- **Parsing**: A suitable parser which can parse all possible expressions created by the grammar rules and composed of all kinds of words and phrases in English. As of our system we select AGFL(Koster, 1991; available at http://www.cs.kun.nl/agfl) as our parser for English grammar because its lexicon can be flexibly extended, its grammar can be readily written and its parsing output can be so arranged that the object-oriented representation of the grammatical elements is easier. We call this kind of parsing result NLML (Natural Language Markup Language) because it uses the style of XML. For example, for the sentence "*I give you a book today.*", the direct parsing result and the **NLML** are shown in Table 1.

- **Representation**: A proper mechanism to transfer the parsing result to the objects representing the grammar elements in the rules. We use Java, the typical **OOP** (Object-Oriented Program) language, to represent the grammatical elements in objects. This technique is called **NLOMJ** (Natural Language Object Modal in Java). The typical objects for describing the English grammar and their relations are shown in Figure 3 with the legend → represents for "use" and --|> represents for "inheritance". The grammatical meanings of the notations of the objects in the figure will be introduced later.

- **Storing**: saving the representation objects in the database. We use MYSQL (available at http://www.mysql.com), a typical freely distributed database system, to save the NLOMJs in the database, which we call **NLDB** (Natural Language Database).

| Direct parsing result | NLML |
|---|---|
| Segment | <mood>statement</mood> |
| statement | <complexity>simple</complexity> |
| simple statement | <subject> |
| simple complete statement without it noun clause | <noun> |
| opt circumstances | <type>perspronoun</type> |
| circumstances | <word>I</word> |
| simple SVOC phrase | <numb>sing</numb> |
| subject(sing, first) | <pers>first</pers> |
| noun phrase(sing, first, nom) | <case>nom</case> |
| noun part(sing, first, nom) | </noun> |
| personal pronoun(sing, first, nom) | </subject> |
| LEX_PERSPRON(sing, first, nom) | <voc> |
| PERSPRON(sing, first, nom) | <verb_type>verb_IO_DO</verb_type> |
| "I" | <tense>present</tense> |
| VOC phrase(sing, first) | <numb>sing</numb> |
| simple VOC phrase(sing, first) | <pers>first</pers> |
| all VOC phrase(sing, first, present) | <verb_word>give</verb_word> |
| real all VOC phrase(sing, first, present) | <circum></circum> |
| verb group(present, ditr, none, none\|to, sing, first) | <circum></circum> |
| verb group without modal(present, ditr, none, none\|to, sing,first) | <indirect_object> |
| opt adverbs | <noun> |
| verb form(present, ditr, none, none\|to, sing, first) | <type>perspronoun</type> |
| LEX_VERBI(none\|to, ditr) | <word>you</word> |
| VERBI(none\|to, ditr) | <numb>NUMB</numb> |
| "give" | <pers>secnd</pers> |
| opt circumstances | <case>dat</case> |
| circumstances | </noun> |
| indirect object phrase | </indirect_object> |
| noun phrase(NUMB, secnd, dat) | <direct_object> |
| noun part(NUMB, secnd, dat) | <prem> |
| personal pronoun(NUMB, secnd, dat) | <type>art</type> |
| LEX_PERSPRON(NUMB, secnd, dat) | <word>a</word> |
| PERSPRON(NUMB, secnd, dat) | </prem> |
| "you" | <noun> |
| direct object phrase | <word>book</word> |
| object phrase | <numb>sing</numb> |
| noun phrase(sing, third, dat) | <type>noun</type> |
| noun part(sing, third, dat) | </noun> |
| normal noun part(sing) | </direct_object> |
| premodifier(sing) | <circum> |
| simple premodifier(sing) | <adv> |
| LEX_ART(sing) | <type>time<type> |
| ART(sing) | <word>today</word> |
| "a" | </adv> |
| rest premodifier | </circum> |
| adj rest premodifier | </voc> |
| noun rest premodifier | |
| real noun(sing) | |
| LEX_NOUN(sing) | |
| NOUN(sing) | |
| "book" | |
| opt circumstances | |
| circumstances | |
| circumstance | |
| adverb | |
| LEX_ADVB | |
| TIMEADVB | |
| "today" | |
| circumstances | |

Table 1 The direct parsing result vs. the NLML

- **World modal**: the common sense knowledge about the world which is embedded in the system. We use Wordnet (available at http://www.cogsci.princeton.edu/~wn/) as a semantic network to retrieve the relationship between the nouns, verbs, adjectives, and adverbs. We use CYC-project (available at http://www.cyc.com) to retrieve the common sense knowledge and then to construct a world modal in common sense.

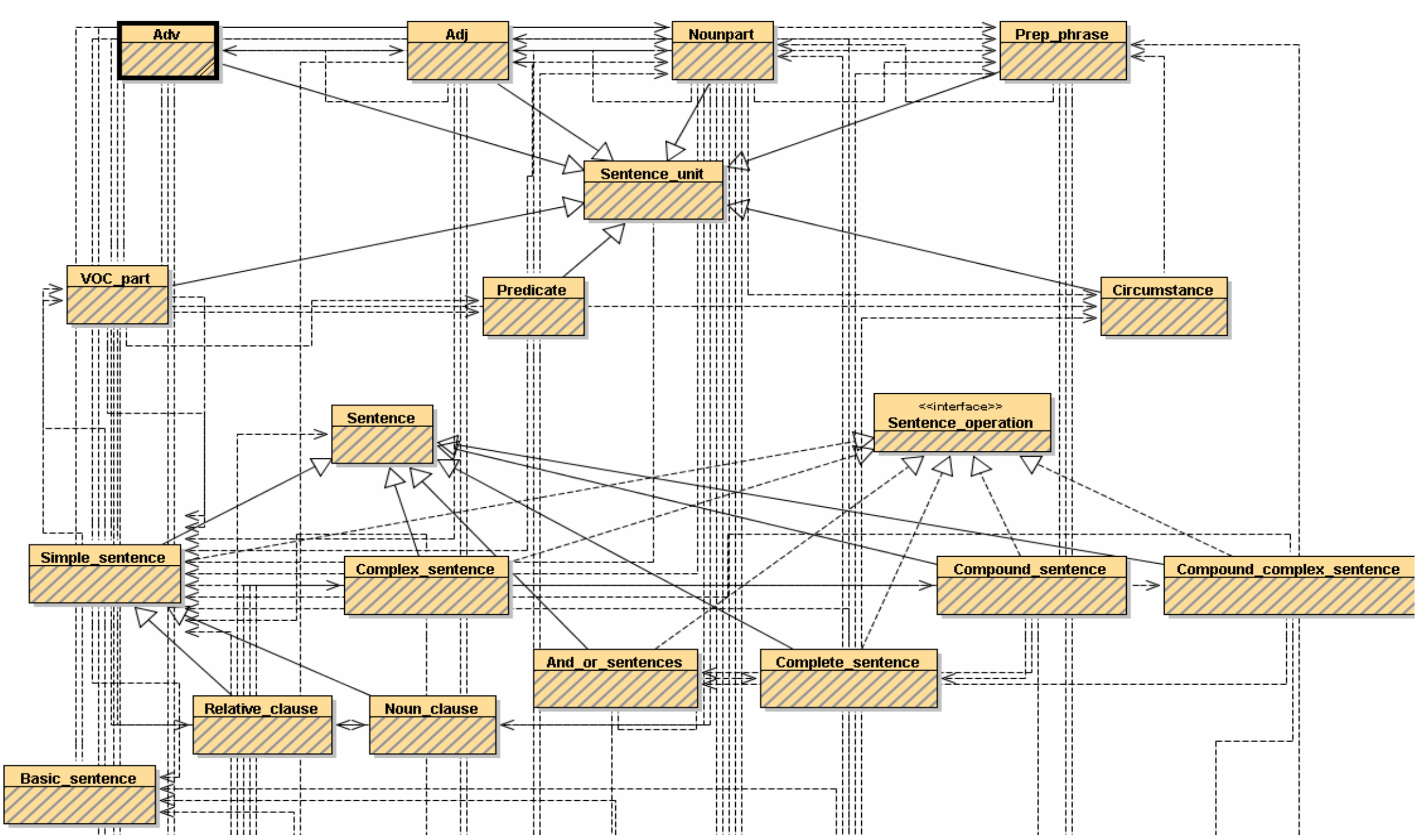

Fig. 3 The objects representing the English grammatical elements and their relations

- **Characterization**: the personality or the character which can be selected by the different users. For example some users prefer one partner who can quietly listen to them; on the contrary some others wish the chatting partner telling them more news or stories.

- **Response**: according to a given expression from the user, deciding whether if it is an idiom which can be directly responded with the muster-matching mechanism or a proper response should be made considering the syntactic semantic factors of the expression which are given in NLOMJ, the context of the dialog existing in the NLDB and the character of the chatbots. We call this mechanism CR (Communicational Response).

  For example if one user inputs "*I give you a book today.*" The CSIEC may response with a question such as "*why do you give me a book?*" or "*Which book will you give me?*" if the CSIEC is curious, and may response with a statement such as "*A book is a written work or composition that has been published (printed on pages bound together).*" If the CSIEC is narrative.

  The great difference between the CR and the muster-matching mechanism is that the response from CR is dynamically generated according to the input sentence and the dialog context, but the response of the muster-matching mechanism is saved in a given database and mostly irrelevant with the dialog content and context, as our previous empirical findings have shown(Jia, 2003).

We call the entire system consisting of the above six components **CSIEC** (Computer Simulator in Educational Communication). The CSIEC is implemented in Java and embedded in a HTTP-server program. The output can not only be text, but also be speech synthesis using Microsoft Agent technology which is accessible at http://www.microsoft.com/msagent. In order to watch and hear the agent character in the local client machine the user should download and install the character and then use the Internet Explorer to browse the website.

Putting all the components together we can see the whole system architecture illustrated in Figure 4.

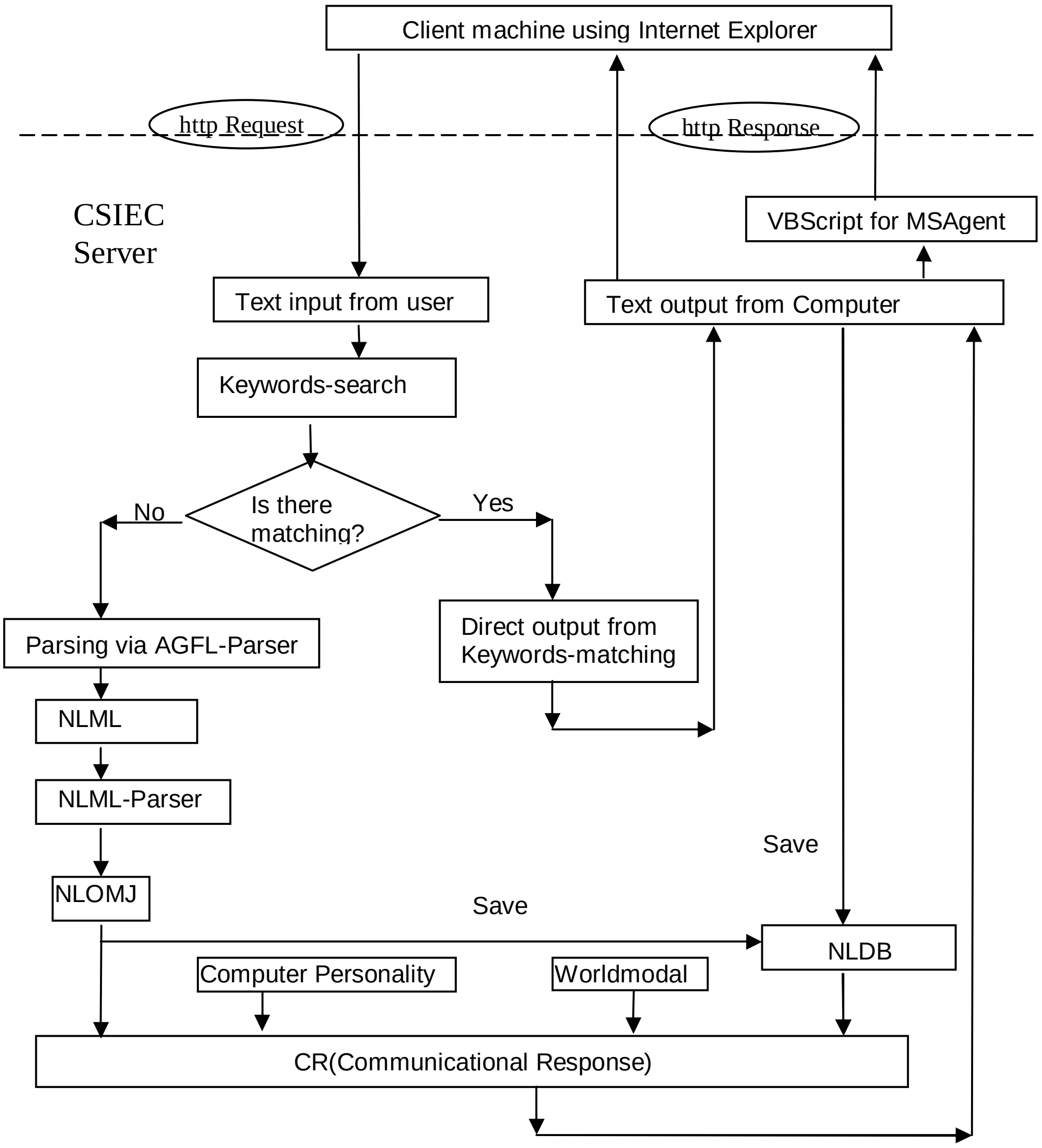

Fig. 4 System architecture of CSIEC

## English Grammar

We use the English grammar from the textbooks of Chalker (Chalker, 1984) and Hanks & Grandison (Hanks & Grandison, 1994). As for the limitation of the pages we can't introduce all the grammar elements in details, but only the important ones which are cited in Figure 3 and their corresponding concrete examples. The notations used to describe the grammar by the author may not be the most appropriate.

- The CSIEC system deals with all kinds of **sentence**s and **phrases**. The sentences types classified according to the **complexity** and their representing Objects in NLOMJ are shown in Table 2. These objects are all subclass of the object "**Sentence"** and all implements the Interface "**Sentence_operation**" defining the standard methods. The compound complex sentence, compound sentence and complex sentence consist of simple sentences connected by conjunctions and/or comma. Therefore the simple sentence is the most element type of sentences.

| Type | Example | Object in NLOMJ |
|---|---|---|
| Compound complex sentence | If it rains today, you will not go, and I will not come. | Compound_complex_sentence |
| Compound sentence | Today you come, he goes, and I wait.<br>Neither you come, nor do I go. | Compound_sentence |
| Complex sentence | If you come I will go. | Complex_sentence |
| Simple sentence | I come today.<br>Both you and I come today.<br>I come and do my job today.<br>This is the book I will give you.<br>What I don't know is how to do this homework. | Simple_sentence |

Table 2 Sentence types according to their complexity

- Every sentence has its **mood**. So the sentences can also be classified by their moods. But we use only the attribute "mood" in the Class "**Sentence**" to represent this type.

| Mood | Example |
|---|---|
| Statement (declarative) | If it rains today, you will not go, and I will not come. |
| Question (interrogative) | What will you do if it rains today? |
| Order (imperative) | Please do your homework if it rains today. |
| Exclamation (exclamative) | What a rainy day! |

Table 3 Sentence types according to their mood

- Every sentence except exclamation has its voice. So the sentences can also be classified by their voice. But we use only the attribute "voice" in the Class "**Sentence**" to represent this type.

| Voice | Example |
|---|---|
| Active | If it rains today, you will not go, and I will not come. I have given him a book.<br>I saw him do his job quickly. |
| Passive | A book has been given to him.<br>He was seen to do his job quickly.<br>It is said that he will come today. |

Table 4 Sentence types according to their voice

- A simple sentence consists of phrases. The main phrases and their corresponding objects are shown in the table 5. All these objects are subclass of the Class "**Sentence_unit**" as they have some common attributes and manipulation methods such as get_text(), toString(), etc.

| Phrase type | Example | Object in NLOMJ |
|---|---|---|
| Adjective phrase | Good, very good, good at physics<br>Good at physics enough to do the job<br>better than that book<br>Adjectives connected by coordinators | Adj |
| Adverb phrase | Fast, quickly, today<br>more quickly than he<br>Too slowly to catch up with him<br>Adverbs connected by coordinators | Adv |
| Preposition phrase | At home<br>In the classroom in the school<br>to that question he has raised<br>Two years ago | Prep_phrase |
| Noun phrase | Real noun: book, person<br>Pronoun: he, she, it, mine, his<br>Noun phrases connected by coordinators<br>May have relative clause | Nounpart |
| Circumstance | Adverb<br>Preposition phrase<br>Participle | Circumstance |
| Predicate | Noun phrase<br>Adjective phrase<br>Preposition phrase<br>Noun clause | Predicate |

Table 5 Phrase types and their corresponding objects in NLOMJ

- **Relative clause** and **noun clause** including gerund clause, participle, infinitive clause and nominal clause are special simple sentence. So they are represented as subclass of the class "**Simple_sentence**".

- A **basic sentence** has only one subject and one VOC part. So a simple sentence is either a basic sentence or a combination of basic sentences connected by conjunctions. The subject is a noun phrase which is an instance of the class Nounpart. The VOC part is represented by an instance of the class VOC_part.

- The **Voc_part** consists of the key verb (and sometimes with auxiliaries) and its corresponding objects, complements or predicates. The verb has different tense: present, past, present perfect, past perfect, present continuous, past continuous, future, past future, present perfect continuous, past perfect continuous, and modal verb (auxiliaries) plus lexical verb with different kinds of tense. The different verb types and their corresponding attachments considered in CSIEC are: be + predicate, copula verb + predicate, verb + indirect object + direct object, verb + direct object, intransitive verb, verb + particle + preposition phrase, verb + preposition phrase, verb + noun phrase + bare infinitive clause, verb + noun phrase +to-infinitive clause, verb + noun phrase + gerund clause, verb + noun phrase + present participle, verb + noun phrase + past participle, verb + noun phrase + predicate, verb + infinitive, verb + participle. The direct object and indirect object are also noun phrases which can be represented by Nounpart.

## References

- Allen, J. (1995). *Natural language understanding*. Benjamin/Cummins Publishing Inc.
- Chalker, Sylvia. (1984). *Current English grammar.* Macmillan Publishers Ltd.
- Chomsky, N. (1956). Three modals for the description of language. IRE transactions PGIT, 2, p.113-124.
- Chomsky, N. (1965). *Aspects of the theory of syntax*. *Cambridge*. MIT Press.
- Chomsky, N. (1969). *Topics in the theory of generative grammar*. Mouton & Co. N.V. Publishers.
- Chomsky, N. (1988). *Language and problems of knowledge*. MIT Press.
- Dale, R., etc. (Ed.). (2000). *Handbook of natural language processing*. Marcel Dekker Inc.
- Hanks, P. & Grandison, A. (1994). *Collins gem English grammar.* HarperCollins Publishers.
- Iwanska, L. M., etc. (2000). *Natural language processing and knowledge representation: language for knowledge and knowledge for language*. AAAI Press/MIT Press.
- Jia, J. (2003). The study of the application of a keywords-based chatbot system on the teaching of foreign languages. Research report at University of Augsburg. Also available at: http://arxiv.org/abs/cs.CY/0310018.
- Jia, J. (2004). The study of the application of a web-based chatbot system on the teaching of foreign languages. In: *Proceedings of the SITE2004 (the 15th annual conference of the Society for Information Technology and Teacher Education).* Accepted.
- Kay, M. (1973). The mind system. in Rustin, R.(eds.): *Natural language processing*. New York: Algorithmics Press.
- Koster, C.H.A. (1991). Affix grammars for natural languages. In: *Lecture Notes in Computer Science*, Vol. 545. Springer-Verlag.
- Pereira, F.C.N. & Warren, D.H.D. (1980). Definite clause grammars for language analysis-A survey of the formalism and a comparison with augmented transition networks. *Artificial Intelligence*, Vol. 13, 3, p.231-278.
- Tapanainen, P. & Järvinen, T. (1997). A non-projective dependency parser. In: *Proceedings of the 5th conference on applied natural language processing (ANLP'97),* p.64-71. ACL, Washington, D.C.
- Weizenbaum, J. (1966). ELIZA--A computer program for the study of natural language communication between man and machine, *Communications of the ACM*, Vol. 9, No. 1, p.36-45.
- Winograd, T. (1972). Understanding natural language. Edinburgh University Press.
- Woods, W.A. (1970). Transition network grammars for natural language analysis. *Communication of the ACM* 13, p.591-606.
- Woods, W.A. (1973). An experimental parsing system for transition network grammars. in R. Rustin(eds.): *Natural language processing*. New York: Algorithmics Press.